\documentclass[final,times]{elsarticle}

\usepackage{lineno,hyperref}
\usepackage[greek,english]{babel}
\modulolinenumbers[5]

\newcommand{\revision}{}

\journal{Acta Astronautica}

\begin{document}

\begin{frontmatter}

\title{Searching for technosignatures in exoplanetary systems with current and future missions}

\author{Jacob Haqq-Misra}
\address{Blue Marble Space Institute of Science, Seattle, WA, USA}

\author{Edward W. Schwieterman}
\address{Department of Earth and Planetary Sciences, University of California, Riverside, CA, USA\\
Blue Marble Space Institute of Science, Seattle, WA, USA}

\author{Hector Socas-Navarro}
\address{Instituto de Astrof\' \i sica de Canarias, C/V\' \i a L\'actea s/n, E-38205 La Laguna, Tenerife, Spain\\
Departamento de Astrof\' \i sica, Universidad de La Laguna, E-38206 La Laguna, Tenerife, Spain}

\author{Ravi Kopparapu}
\address{NASA Goddard Space Flight Center, 8800 Greenbelt Road, Greenbelt, MD 20771, USA}

\author{Daniel Angerhausen}
\address{Blue Marble Space Institute of Science, Seattle, WA, USA\\
ETH Z\"urich, Institute for Particle Physics and Astrophysics, Switzerland}

\author{Thomas G. Beatty}
\address{Department of Astronomy and Steward Observatory, University of Arizona, Tucson, AZ 85721, USA}

\author{Svetlana Berdyugina}
\address{Leibniz-Institut für Sonnenphysik}

\author{Ryan Felton}
\address{Catholic University of America, Washington D.C., USA\\
NASA Ames Research Center, Mountain View, CA}

\author{Siddhant Sharma}
\address{Blue Marble Space Institute of Science, Seattle, WA, USA}

\author{Gabriel G. De la Torre}
\address{Neuropsychology and Experimental Psychology Lab. University of Cadiz. Spain}

\author{D\'aniel Apai}
\address{Department of Astronomy, The University of Arizona, Tucson, AZ 85721, USA\\
Lunar and Planetary Laboratory, The University of Arizona, Tucson, AZ 85721, USA}

\author{and the TechnoClimes 2020 workshop participants}

\begin{abstract}
Technosignatures refer to observational manifestations of technology that could be detected through astronomical means. Most previous searches for technosignatures have focused on searches for radio signals, but many current and future observing facilities could also constrain the prevalence of some non-radio technosignatures. This search could thus benefit from broader participation by the astronomical community, as contributions to technosignature science can also take the form of negative results that provide statistically meaningful quantitative upper limits on the presence of a signal. This paper provides a synthesis of the recommendations of the 2020 TechnoClimes workshop, which was an online event intended to develop a research agenda to prioritize and guide future theoretical and observational studies technosignatures. The paper provides a high-level overview of the use of current and future missions to detect exoplanetary technosignatures at ultraviolet, optical, or infrared wavelengths, which specifically focuses on the detectability of atmospheric technosignatures, artificial surface modifications, optical beacons, space engineering and megastructures, and interstellar flight. This overview does not derive any new quantitative detection limits but is intended to provide additional science justification for the use of current and planned observing facilities as well as to inspire astronomers conducting such observations to consider the relevance of their ongoing observations to technosignature science. This synthesis also identifies possible technology gaps with the ability of current and planned missions to search for technosignatures, which suggests the need to consider technosignature science cases in the design of future mission concepts.
\end{abstract}

\end{frontmatter}

\section{Introduction} 
\label{sec:intro}

The detection of exoplanets with space- and ground-based telescopes has suggested that most stars in the galaxy host planets \citep{cassan2012one, dressing2015occurrence}. A recent statistical analysis by \citet{bryson2021} examined the Kepler planet catalog and found that about half of Sun-like stars should host a terrestrial planet within the liquid water habitable zone, so that there should be about four nearby habitable terrestrial planets, on average, among the G- and K-dwarf systems within 10\,pc from Earth. Similar approaches to calculating this occurrence rate have found that temperate terrestrial planets are common around M dwarf stars \citep{Thompson2018, Hardegree-Ullman2019,Bryson2020note}. Specifically, \citet{dressing2015occurrence} found that about 1 in 6 M-dwarf systems should host an Earth-sized planet in the habitable zone. One of the goals of exoplanet science is the spectral characterization of exoplanetary atmospheres and systems, and the expected prevalence of habitable planets serves to motivate the search for spectroscopic evidence of life in such systems. While statistical estimates for the prevalence of potentially habitable planets vary significantly, most studies suggest they are numerous enough to motivate this search. 

\begin{figure*}[ht!]
\centerline{\includegraphics[width=5in]{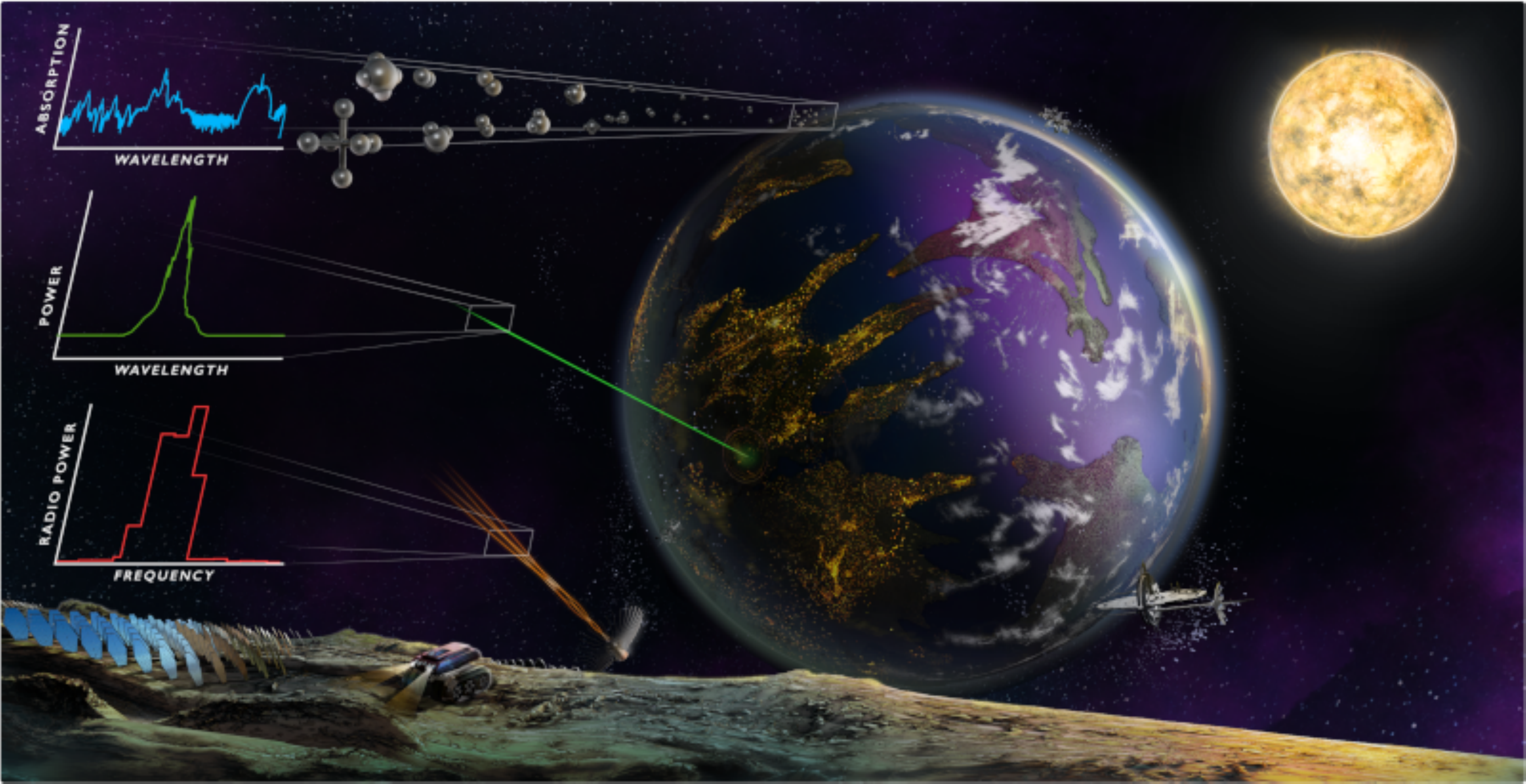}}
\caption{A concept image illustrating various types of technosignatures described in this paper, including atmospheric, optical, and radio technosignatures. \revision{Atmospheric technosignatures may include obviously artificial molecules such as sulfur hexafluoride (SF$_{6}$) in addition to common molecules expected for an inhabited terrestrial planet, such as oxygen (O$_{2}$), carbon dioxide (CO$_{2}$), and methane (CH$_{4}$). The top left inset shows the absorption cross-sections of S$_{6}$ \citep{sharpe2004gas}. Optical technsignatures include highly collimated  laser pulses that can outshine the host star at narrow wavelengths (i.e., Optical SETI). The middle left inset illustrates the narrow power distribution of an optical (green) laser pulse. Active radio beacons or passive radio leakage from the planetary surface, orbit, or elsewhere in the stellar system would be recognizably artificial (i.e., traditional SETI). The bottom left inset illustrates the narrow distribution of power versus frequency anticipated for an artificial radio signal. Additional potentially detectable technosignatures in this planetary system include artificial lighting on the planetary nightside \citep[e.g.,][]{beatty2021}, recognizable spectral breaks from solar arrays  on the planet's moon \citep[e.g.,][]{Lingam2017}, and anomalous transit signatures from the orbiting habitats and satellite arrays \citep[e.g.,][also see section \ref{sec:megastructures}]{socas2018possible}}. 
\label{fig:technosigconcept}}
\end{figure*}

Spectroscopic characterization of exoplanetary atmospheres has already been demonstrated for gas giants \citep[e.g.][]{dave2002, sing2016} and mini-Neptunes \citep[][]{paper12019,paper22019}. This will eventually become possible for terrestrial planets with the next generation of telescopes \citep[][]{luvoir2019luvoir,Gaudi2020}. The idea that disequilibrium chemistry in a planetary atmosphere could be linked to the presence of a biosphere was first proposed by \citet{lovelock1975thermodynamics} and inspired further discussions of plausible ``biosignatures'' on exoplanets that would indicate the presence of life. Even prior to any observations of terrestrial exoplanet spectra, the astrobiology community has already developed theoretical frameworks for evaluating possible spectral biosignatures (as well as false positives) that could be detected by upcoming missions \citep[e.g.][]{shawn2011, seager2012astrophysical, kaltenegger2017characterize, grenfell2017review,Schwieterman2018, Meadows2018, Arney2018, Catling2018, Walker2018, Fujii2018, o2019expanding, lammer2019role}. These ongoing efforts seek to understand the observational prospects for biosignatures and develop a library of possibilities that might be detected to help prepare for the future discovery of an exoplanetary biosphere.

A logical extension to the search for extraterrestrial life through biosignatures is the search for evidence of extraterrestrial technology. The idea of searching for ``technosignatures'' has been considered by astronomers for more than half a century, with initial efforts focused on the possibility of detecting extraterrestrial radio transmissions \citep{cocconi1959searching,drake1961project}. Many other possible technosignatures have been suggested with observable properties at ultraviolet, visible, and infrared regions of the spectrum, which include waste heat \citep{dyson60,GHAT2,kuhn2015global,carrigan09b}, energy-intensive illumination \citep{Schneider2010,Loeb11,Kipping16}, surface modifications \citep{Lingam2017}, atmospheric pollution \citep{Schneider2010,Lin2014,Stevens16,Kopparapu2021}, stellar pollution \citep{shklovskii1966,Whitmire80,Stevens16}, non-terrestrial artifacts \citep{bracewell1960communications,freitas1980search,Rose2004,JR2012}, and megastructures \citep{dyson60,arnold2005transit,Forgan13,GHAT4}. The range of possible technosignatures draws from the observable evidence of technology on Earth today as well as future projections of technology that are known to be possible. The search for technosignatures can complement the search for biosignatures by looking for spectral evidence of technology in a planet's atmosphere or within planetary systems (Fig \ref{fig:technosigconcept} provides a conceptual illustration of several example technosignatures); however, the theoretical basis for understanding and prioritizing possible technosignatures remains in a state of infancy compared to biosignature science  \citep{wright2019,jacob2020}.

Funding remains a limiting factor in advancing technosignature science, but recent years have shown a renewed interest in technosignatures by public and private funding agencies, motivated in part by advances in exoplanet detection and characterization. The most comprehensive search for technosignatures is the privately-funded Breakthrough Listen initiative, which began in 2015 as a 10-year \$100M effort to conduct radio and optical observations of over a million stars \citep{worden2017breakthrough,isaacson2017breakthrough}. \revision{A narrowband signal of interest, BLC-1, was detected by Breakthrough Listen using the Parkes radio telescope in the direction of Proxima Centauri \citep{smith2021radio}; later analysis found BLC-1 to be a product of local interference, but the signal provided an opportunity to test Breakthrough Listen's analysis pipeline and verification framework \citep{sheikh2021analysis}.} \revision{Other ongoing efforts include} observations by the SETI Institute's Allen Telescope Array \citep{tarter2011first,harp2016seti} as well as numerous other efforts by research groups around the world searching for radio technosignatures \citep[e.g.,][]{montebugnoli2010seti,rampadarath2012first,zhang2020first}, which are supported by combinations of public and private investments as well as through commensal observing strategies. 

Renewed interest in technosignatures by NASA began in 2018, which led the agency to organize a workshop (``NASA technosignatures workshop,''\footnote{\url{https://www.hou.usra.edu/meetings/technosignatures2018/}} held in September 2018 at the Lunar and Planetary Institute, Houston, USA) to inform the agency about its potential role in the search for technosignatures. The resulting report summarized the four objectives of the workshop, which included defining the state of the field, identifying opportunities for near-term advances, understanding the potential for future advances, and advising NASA on the role of partnerships in technosignature science \citep{2018arXiv181208681N}. The 2018 workshop report remains a valuable resource for investigators who are interested in learning more about the history and status of technosignature science, which includes a discussion of previously conducted searches. A comprehensive curated SETI \revision{bibliography} is also maintained as a searchable library \citep{lafond2021furthering}.
\footnote{The SETI Institute also maintains a database of published searches for technosignatures: \url{https://technosearch.seti.org/}} 

A second workshop was sponsored by NASA in August 2020 (“TechnoClimes 2020,”\footnote{\url{https://technoclimes.org/}} Blue Marble Space Institute of Science), with the goal of developing a research agenda for technosignature science. TechnoClimes was organized as a 5-day online event to accommodate participation during the COVID-19 pandemic, with a total of 53 participants from 13 countries. This paper represents one of four objectives of the TechnoClimes research agenda for technosignatures. The objective of this paper is to encourage a broader range of astronomers to consider the relevance of technosignatures to their research by serving as a resource that describes the detectability of various non-radio technosignatures with current and future missions. The second research agenda objective was to develop mission concepts that are optimized for technosignature science and was explored by \citet{socas2021concepts}, which included overview of future mission possibilities as well as a metric for ranking the relative detectability of technosignatures. The third research agenda objective recognized that the first detection of a technosignature might be an anomalous finding in a field other than astrobiology, which led \citet{singam2020evaluation} to develop a conceptual framework for evaluating ``non-canonical astrophysical phenomena'' as potential technosignatures. The final objective identified by the workshop is the need to build the technosignatures community, which includes broadening international participation as well as expanding support for early career researchers; these recommendations and others discussed at the workshop were summarized for the Planetary Science Decadal Survey by \citet{Sheikh2020decadal}.

This paper specifically focuses on the possibility of detecting technosignatures in exoplanetary systems with current and future missions and observatories. This approach is intended to highlight the potential capabilities of available technology to search for technosignatures, even if technosignature science is ancillary to the primary mission objectives. It is important to emphasize that negative results still provide constraints on the prevelance of technosignatures, so the lack of a positive technosignature detection should not be conflated with a lack of progress in technosignature science \citep{wright2019}. 

The focus on exoplanetary systems is not intended to neglect any technosignatures that could be present within the solar system, such as the possibility of extraterrestrial artifacts on the surfaces of planets or at stable Lagrange points \citep{bracewell1960communications,freitas1983search,benford2019looking,benford2020a,benford2020b}. The search for solar system technosignatures was a focus of discussion during TechnoClimes, which is relevant to a wide range of planetary science missions that are exploring Venus, Mars, Titan, Enceladus, asteroids and other objects. For example, a search for technosignatures such as non-terrestrial artifacts \citep[e.g.][]{haqq2012likelihood,davies2013searching} could be conducted on the moon by analyzing high-resolution images taken by the Lunar Reconnisance Orbiter \citep{moseley2019unsupervised}. Full discussion of the use of Solar System exploration missions to constrain the prevalence of technosignatures is beyond the scope of this paper, but such an analysis would be valuable future work. We emphasize that thinking about technosignatures is relevant to solar system science as well as to exoplanets. 

\revision{Finally, the focus of this paper is on concepts for detecting technosignatures in exoplanetary systems using non-radio methods. Our exclusion of detailed analysis of radio technosignatures is this paper is not at all intended to diminish the value of radio searches in technosignature science, and we recognize that much work remains to be done to better constrain the prevalence of radio technosignatures \citep{wright2018much}. Instead, our choice to focus on non-radio technosigantures intended to highlight the numerous other methods of searching for technosignatures that could be relevant to missions and facilities that have not traditionally considered technosignature science. Radio astronomy has a longer history of considering technosignature science, and upcoming facilities like the Square Kilometer Array even include the search for radio technosignatures through direct and commensal observing among its science objectives \citep{siemion2014searching,bock2015concept}. The need to understand and minimize radio interference is also a problem for general radio astronomy that also relates to radio technosignatures in particular, and ongoing efforts by the scientific community are even attempting to protect regions of the lunar farside from such interference \citep{maccone2008protected,maccone2019moon}. Such efforts are examples of the ongoing role of technosignature science within existing radio astronomy. Our goal for this paper is to inspire other disciplines within astronomy to similarly consider the relevance of technosignature science to their existing and planned facilities.}

This paper begins with a brief description of future and current missions that could search for non-radio technosignatures in section \ref{sec:missions}. \revision{This brief discussion of missions and facilities relevant to technosignatures allows us to refer to specific concepts and architectures in the following discussion of technosignature classes.} We discuss the capabilities of such missions for detecting atmospheric technosignatures in section \ref{sec:planetary}, surface technosignatures in section \ref{sec:surface}, \revision{optical beacons in section \ref{sec:optical}}, megastructures in section \ref{sec:megastructures}, and interstellar flight in section \ref{sec:spaceflight}. This presentation is intended to summarize current and future capabilities as well as serve as a reference for investigators interested in expanding their existing research to include technosignatures. None of the mission concepts considered in this paper are capable of searching for all identifiable technosignatures, so the search for technosignatures will increase its likelihood of success through commensal and direct observing programs across a broad range of missions.

\section{Current and future missions \revision{and facilities}}\label{sec:missions}

\begin{figure*}[ht!]
\centerline{\includegraphics[width=5in]{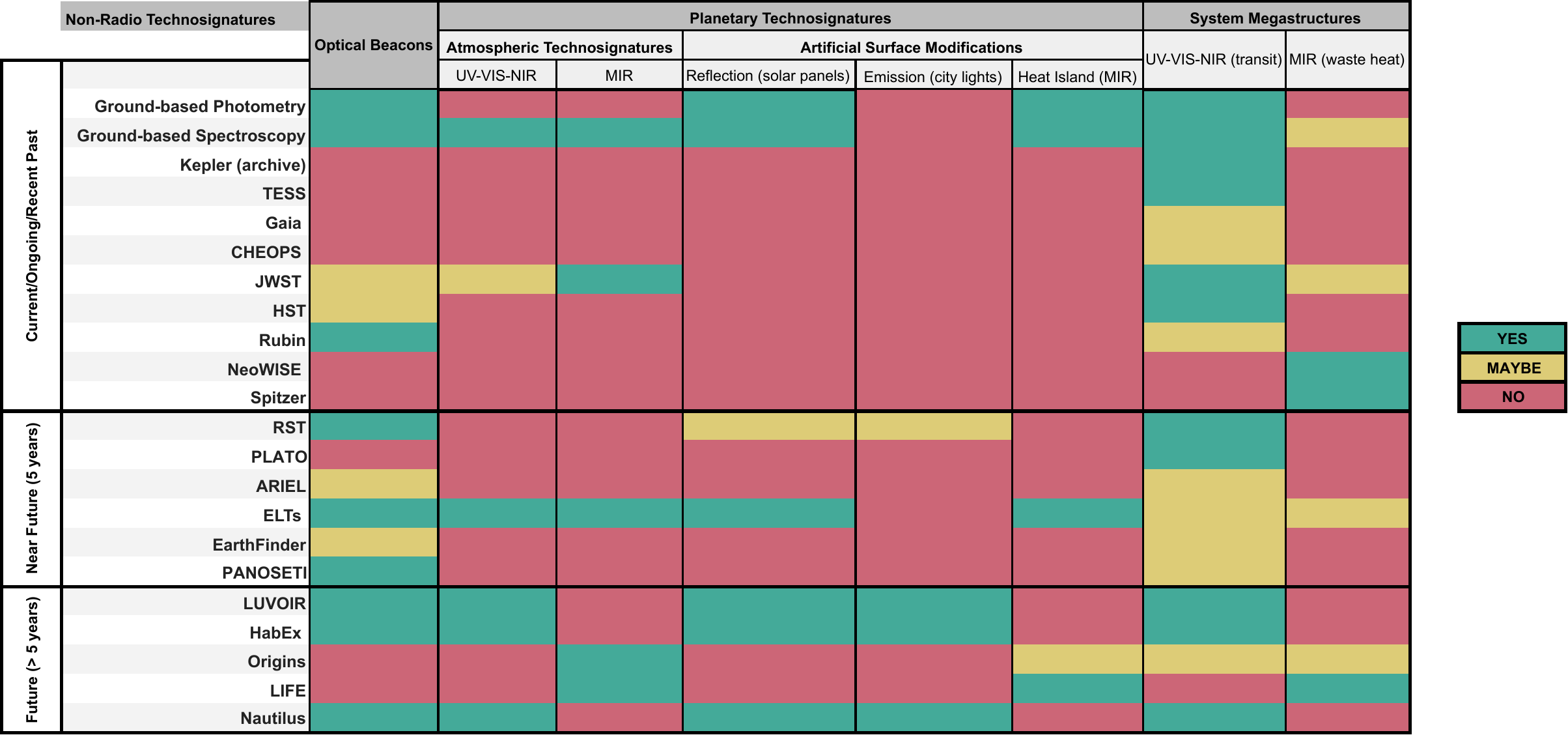}}
\caption{Capabilities for detecting technosignatures with recent, ongoing, and future missions \revision{and facilities}. Cells colored green indicate there is at least one stellar system where the given technosignature could be detectable with the mission or observatory. A green cell further indicates there is a peer-reviewed publication that has evaluated the hypothetical detectability of that technosignature. A yellow cell indicates the potential detectability of that technosignature in at least one stellar system, but that further study is needed. A red cell indicates that the given technosignature is not detectable with that observatory or mission architecture for any stellar systems. \revision{Note that we include all ground-based instrumentation in the ground-based photometry and ground-based spectroscopy categories, although specific observatory-instrument combinations may only access a subset of indicated technosigatures. For example, the Gemini Planet Imager (GPI) could plausibly detect optical beacons \citep{vides2019model}, but no other indicated technosignatures. See text accompanying sections \ref{sec:planetary}, \ref{sec:surface}, \ref{sec:optical}, and \ref{sec:megastructures} for more specific examples of current, future, or potential ground-based facilities capable of detecting the indicated technosignatures}. Examples are meant to be illustrative rather than exhaustive, and focus on missions or facilities with the capability of characterizing terrestrial planets. 
\label{fig:currentfuturemissions}}
\end{figure*}

\revision{We begin our analysis with} a brief overview of the current, near future, and distant future missions that could collect data to constrain the prevalence of technosignatures in exoplanetary systems. This list of missions corresponds to the first column of Fig. \ref{fig:currentfuturemissions}, with rows showing the capabilities of searching for different technosignatures with each mission. \revision{The discussion in this section is intended to provide context for specific concepts and architectures that will be referenced in the following sections. Readers that are already familiar with these mission concepts, or who are using this paper as a reference, may choose to advance to the next section and consult section \ref{sec:missions} when needed as a reference.}

\subsection{Current, ongoing, recent, and past missions}

Current, ongoing, recent, and past missions already have data available for analysis, many of which could be used to constrain the search for technosignatures. Such missions represent opportunities to develop new observing campaigns for technosignatures in dedicated or commensal modes (for missions still operating) as well as new data mining techniques (for completed missions). 

A range of \textit{ground-based} facilities can detect exoplanet transits through \textit{photometry} \citep[e.g.][]{deeg2009ground,nascimbeni2013blue} and characterize exoplanet atmospheres \textit{spectroscopy} \citep{bean2010ground,diamond2018ground} at ultraviolet, visible, and infrared wavelengths. Such facilities can also be used to corroborate observations by space telescopes and further constrain the properties of known exoplanet systems.

The \textit{Kepler} space telescope and follow-on mission K2, when the telescope lost two of its reaction wheels, were exoplanet transit detection missions utilizing a 0.95 m aperture to scan approximately 200,000 stars. With the decommission of \textit{Kepler} and K2 in September 2018, all of the astronomical data (i.e. light-curves) is now archived in the Mikulski Archive for Space Telescopes, which allows scientists searchable access to the Kepler/K2 mission data.\footnote{https://archive.stsci.edu/}

The Transiting Exoplanet Survey Satellite (\textit{TESS}) is an exoplanet transit detection mission that launched in April 2018 and will observe approximately 85\% of the sky. The goal of TESS is to detect small planets in orbit around bright stars ranging from M-dwarfs to F-dwarfs \citep{barclay2018revised}. \revision{The TESS mission has already detected over 5,000 planets and planet candidates and may detect over 12,000 planets during its extended seven year mission duration \citep{Kunimoto2022}}. 

The \textit{Gaia} mission was launched in December 2013 to conduct precise astrometric measurements of about a billion celestial objects \citep{prusti2016gaia}. The catalog of Gaia objects is primarily populated by stars, but the mission can also use astrometry to detect exoplanets \citep{perryman2014astrometric}.

The Characterising Exoplanet Satellite (\textit{CHEOPS}) conducts photometric observations of transiting planets to constrain measurements of planetary radius \citep{deline2020expected}. CHEOPS launched in December 2019 and can improve constraints on observed properties of planets detected by other missions, such as TESS \citep{gaidos2017exoplanet}.

The James Webb Space Telescope (\textit{JWST}) \citep{gardner2006james,Greene2016} is a 6.6 m diameter infrared space based observatory located at the L2 Lagrange point\revision{, which successfully launched at the end of 2021}. The telescope will focus on four main science themes: The End of the Dark Ages: First Light and Reionization, The Assembly of Galaxies, The Birth of Stars and Protoplanetary Systems, and Planetary Systems and the Origins of Life.

The Hubble Space Telescope (\textit{HST}) is a 2.4 m diameter general purpose near-infrared, optical, and ultraviolet observatory that was launched in 1990 and has been sustained through successive repairs. HST has demonstrated the capabilities for direct detection of exoplanets \citep{currie2012direct} as well as the spectroscopic characterization of some exoplanet atmospheres \citep{swain2014detection}.

The Vera C. Rubin Observatory, previously known as the Large Synoptic Survey Telescope, is a ground-based observatory under construction in Chile with a 8.4 m diameter mirror. The Rubin Observatory will conduct regular optical observations of the complete sky, which could be adequate for detecting transits of exoplanets \citep{lund2014transiting}.

The Wide-field Infrared Survey Explorer (\textit{WISE}) is a 0.4 m diameter infrared space telescope that was launched in December 2009 and conducted all-sky surveys until 2011 when it entered hibernation mode. Such observations included the ability to detect debris disks in previously identified exoplanet systems \citep{lawler2012debris}. The mission was re-activated as the Near-Earth Object Wide-field Infrared Survey (\textit{NEOWISE}) in 2013.

The \textit{Spitzer} space telescope was a 0.85 m diameter infrared space telescope that launched in August 2003 and retired in January 2020. Although Spitzer was not designed for expolanet science, it detected the first thermal emission from a hot Jupiter planet and has been used to conduct transit, microlensing, astrometry, and direct imaging searches and characterization \citep[e.g.][]{grillmair2007spitzer,deming2020highlights}.

\subsection{Near-future missions and observatories}

Near-future missions are within approximately 5-10 years of being completed. Such missions represent the emerging state-of-the art capabilities to characterize extrasolar systems, which includes opportunities for constraining the prevalence of technosignatures. However, near-future space-based missions in particular have already gone through the design phase, so there is no longer an opportunity to adjust their capabilities to be more sensitive to potential technosignatures. 

The Nancy Grace Roman Space Telescope (\textit{RST}), previously known as the Wide-Field Infrared Survey Telescope (\textit{WFIRST}), is a 2.4 m diameter wide-field instrument under construction with an optical coronagraph \citep{Spergel2015}. RST is a general purpose astronomical instrument that will include the capability to detect exoplanets through direct observations or microlensing \citep{johnson2020predictions}.

The PLAnetary Transits and Oscillations of stars (\textit{PLATO}) mission is a wide-field instrument with 34 small aperture telescopes that will observe and characterize up to one million stars, which is expected to detect thousands of exoplanets in transit \citep{rauer2014plato}. 

The Atmospheric Remote-sensing Infrared Exoplanet Large-survey (\textit{ARIEL}) mission is a meter-class space telescope that will detect and characterize exoplanets at visible and infrared wavelengths. ARIEL is expected to observe at least 500 exoplanets, with an emphasis on warm planets that are closely orbiting their host star \citep{Tinetti2018}.

Extremely Large Telescopes (\textit{ELTs}) refer to a class of ground-based telescopes with a diameter ranging from 20 to 100 m, which can include ultraviolet, optical, and infrared wavelengths. The high sensitivity enabled by ELTs could allow for detection and characterization of Earth-sized exoplanets \citep{carlomagno2020metis,wagner2021imaging}, especially if combined with a coronagraph \cite[see e.g.,][]{cavarroc2006fundamental,2013ApJ...764..182S, 2017A&A...599A..16L}. The European-Extremely Large Telescope (E-ELT) \citep{gilmozzi2007european}, Giant Magellan Telescope (GMT) \citep{johns2006giant} the Thirty Meter Telescope (TMT) \citep{skidmore2015thirty} are ELTs currently under construction. 

\textit{EarthFinder} is a mission concept study to conduct precise radial velocity measurements of nearby exoplanets using a $\sim$1.5 m diameter telescope, which would include coverage across ultraviolet, optical, and near-infrared wavelengths \citep{2020arXiv200613428P}.


\textit{PANOSETI} is an all-sky survey under design that would search the entire northern hemisphere at fast time resolution ranging from nanoseconds to seconds. PANOSETI would be a dedicated facility for technosignature searches, but observations will also include study of other astrophysical phenomena \citep{2019BAAS...51g.264W, 2020SPIE11445E..8BB}.


\subsection{Distant future missions}

Distant future mission concepts are at least 10-15 years or more away from being constructed, if selected for funding. Such mission concepts are technologically feasible, but actual data from such missions may not be forthcoming for decades. Such missions can serve as a motivation for constraining theoretical efforts to understand the detectability of various technosignatures. For most of these mission concepts, final designs have not yet been decided and there remains an opportunity to influence their ultimate capabilities through prospective modeling and theoretical work. \revision{These mission concepts may also provide inspiration for other future missions to be designed by national or commercial space agencies.}

\textit{LUVOIR} (Large Ultraviolet Optical Infrared Surveyor) \citep{luvoir2019luvoir} was one of four Astro2020 Decadal Survey Mission Concept Studies for a highly capable, multi-wavelength space observatory that would enable great leaps forward in a broad range of science, from the epoch of reionization, through galaxy formation and evolution, star and planet formation, to solar system remote sensing. LUVOIR would also have the major goal of characterizing a wide range of exoplanets, including those that might be habitable\footnote{\url{https://www.luvoirtelescope.org/}}. Two possible LUVOIR architectures were studied, one with a 15-m primary mirror and another with an 8-m primary mirror, both with internal coronagraphs used to null the light from the host star and reveal the planetary contribution. The 2020 Astrophysics Decadal Survey \citep{NationalAcademiesofSciencesEngineering2021} recommended NASA pursue the design and construction of a 6-m class infrared / optical/ ultraviolet surveyor mission with capabilities similar, but not identical, to the 8-m LUVIOR-B architecture described in the LUVOIR report \footnote{\url{https://doi.org/10.17226/26141}}. The Decadal Survey also recommended a Great Observatories program that includes investments in technological development across the electromagnetic spectrum, including the development of far-IR and X-Ray probe missions. Because there remain uncertainties regarding the implementation of the Decadal Survey recommendations, we describe the other relevant flagship concept studies that are  illustrative of potential future technological development. 

\textit{HabEx} \citep{Gaudi2019} was a second mission concept studied for the Astro2020 Decadal Survey. It would  directly image planetary systems around Sun-like stars at ultraviolet, optical, and infrared wavelengths. HabEx would be sensitive to all types of planets; however its main goal would be to directly image Earth-like exoplanets for the fist time, and characterize their atmospheric composition. By measuring the spectra of these planets, HabEx would search for signatures of habitability such as water, and be sensitive to gases in the atmosphere possibly indicative of biological activity, such as oxygen or ozone \footnote{\url{https://www.jpl.nasa.gov/habex/}}. A major distinguishing feature of the proposed HabEx architecture was an external starshade spacecraft to be used for starlight suppression. The Astro2020 survey recommendation lies along the spectrum of options delineated by the HabEx and LUVOIR reports, but does not recommend the starshade technology to be used in conjunction with the infrared / optical / ultraviolet surveyor.    

\textit{Origins} \citep{Cooray2019} was the third mission concept study considered by the 2020 Astrophysics Decadal Survey. It would trace the history of our origins from the time dust and heavy elements permanently altered the cosmic landscape to present-day life. Origins would operate at mid- and far-infrared wavelengths and offer powerful spectroscopic instruments and sensitivity three orders of magnitude better than that of Herschel, the largest telescope flown in space to date\footnote{\url{https://asd.gsfc.nasa.gov/firs/}}. The Astro2020 Decadal Survey included the recommended development of a Far-IR probe, which may have some overlapping capabilities with the Origins mission as studied. 

\textit{Lynx} was the fourth mission concept considered by the 2020 Astrophysics Decadal Survey. It would have studied high-energy astrophysical phenomena in X-rays including black hole accretion disks, galaxy evolution, and the birth and death of stars \footnote{\url{https://wwwastro.msfc.nasa.gov/lynx/}}. Its utility for studying temperate terrestrial planets would be limited, but it would be capable of observing planetary transits in X-ray wavelengths and characterizing host star activity in the X-ray regime. The Astro2020 Decadal Survey included the recommended development of an X-ray probe, which may have some overlapping capabilities with the Lynx mission as studied. Currently few potential X-Ray technosignatures have been described in the literature, but this may be an area of future interest for the field.

\textit{LIFE} (Large Interferometer for Exoplanets) is a project initiated in Europe with the goal to consolidate various efforts and define a roadmap that eventually leads to the launch of a large, space-based mid-infrared nulling interferometer \citep{defrere2018space,quanz2019atmospheric}. Detailed simulations including all astrophysical noise sources show that LIFE (consisting of a 4-telescope array with at least 2m apertures) can detect more than 300 sub-Neptune sized planets \citep{2021arXiv210107500L}, which includes dozenz of rocky and temperate planets \citep[c.f.][]{kammerer2018simulating,quanz2019atmospheric}. In the characterization phase, LIFE will measure mid-infrared spectra for a subset of these worlds, although mission priorities remain to be defined.

The \textit{Nautilus} Space Observatory\footnote{\url{https://nautilus-array.space}} \citep[][]{Apai2019Nautilus,Apai2019SPIE,Apai2019NautilusAstro2020} is a misson concept that aims to spectroscopically survey 1,000 exo-Earth candidates for atmospheric biosignatures using transmission spectroscopy at near-ultravoilet, visible and near-infrared wavelengths. The novel aspects of Nautilus is that it envisions an incoherent array of cost-effectively replicated, identical telescopes that are launched in the same launch vehicle, driving down costs and risks. Replication of space telescopes typically faces a key bottleneck from the cost-effective production of primary mirrors. Nautilus circumvents this bottleneck through the use of Multi-order Diffractive Engineered Material (MODE) lenses \citep[][]{Milster2020}, which provide high optical quality yet very low-mass optical elements that can be replicated efficiently via optical molding \citep[][]{Zhang2020,Casstevens2018}. The Nautilus Space Observatory concept envisions the launch of 35 unit telescopes, each equipped with a 8.5 m diameter MODE lens and a simultaneous visible/near-infrared low-resolution imaging spectrograph. A University of Arizona-based group has demonstrated the core MODE lens technology and is currently building scaled-down prototypes. A pathfinder Probe-class, single-unit version of Nautilus was proposed to the Astro2020 Decadal Survey \cite[][]{Apai2019NautilusAstro2020}.

While many of the mission concepts described above may not continue development after the result of the Astro2020 report, their studied capabilities provide a quantitative reference baseline for potentially future missions with similar architectures. Several existing technosignature and biosignatures studies reference these mission concepts when evaluating the potential detectability of a given signal. Therefore, we continue to reference these concepts when describing specific technosignature types below. 

\section{Atmospheric technosignatures}\label{sec:planetary}

We begin our consideration of possible technosignatures in a planet's atmosphere, which could range from the ultraviolet to mid-infrared region. Analogous to the search for planetary biosignatures, no significant searches for non-radio planetary technosignatures have yet been conducted.

Atmospheric technosignatures are gases that are produced by artificial means either as an incidental byproduct of industrial civilization or for a specific purpose, perhaps to manage planetary climate \citep{marinova2005radiative,schneider2010far}. Examples of atmospheric technosignatures can be derived by studying human civilization’s own atmospheric outputs, which include nitrogen dioxide (NO$_2$), chlorofluorocarbons (CFCs), hydrofluorocarbons (HFCs), perfluorocarbons (PFCs), sulfur hexafluoride (SF$_6$), and nitrogen trifluoride (NF$_3$). This list is far from complete, as the development of a comprehensive accounting of the detectability limits of atmospheric technosignatures remains an active area of research.

\begin{figure*}[ht!]
\centerline{\includegraphics[width=5in]{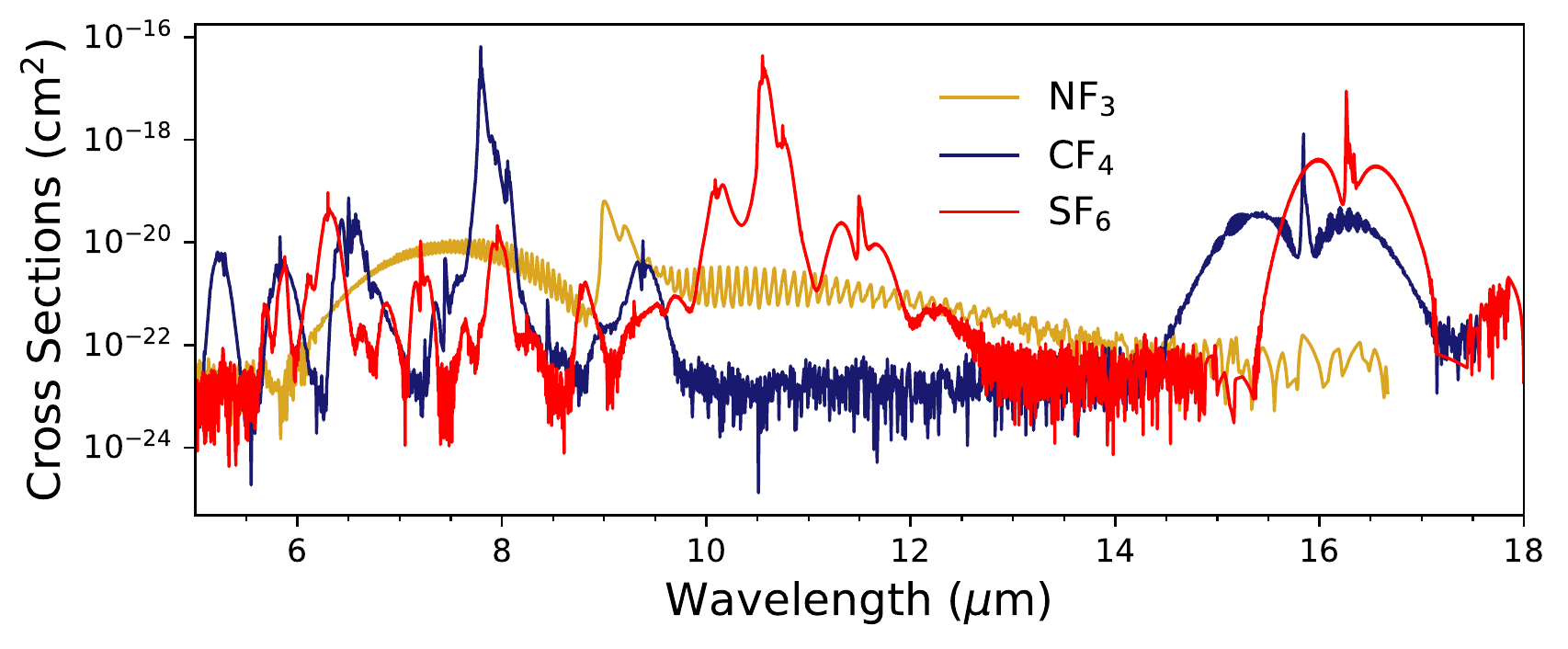}}
\caption{Cross-sections for a subset of potential atmospheric technosignature molecules including NF$_{3}$, CF$_{4}$, and SF$_{6}$. \revision{These long-lived artificial fluorine-containing gases have very limited abiotic sources and may serve as indicators of deliberate climate modification on inhabited terrestrial planets \citep[e.g.,][]{marinova2005radiative}. These molecules could be detected in infrared emission or transmission spectral observations (see section \ref{sec:planetary}).} Cross-section data were sourced from \citet{sharpe2004gas} via \citet{kochanov2019infrared}.
\label{fig:atmtech}}
\end{figure*}

The detectability of potential technosignature gases will be strongly dependent on their intrinsic absorption features (e.g., Fig \ref{fig:atmtech}), their abundance in a planetary atmosphere, the distance to the target system, the characteristics of the target system (e.g., spectral type of host star), and the observing mode, wavelength coverage, spectral resolution, and instrumental noise floor of the observing facility. For example, \citet{lin2014detecting} specifically estimate the detectability of CF$_4$ and CCl$_3$F in the mid-infrared part of the spectrum for an Earth-like planet transiting a white dwarf star and estimate that concentrations of ten times modern concentrations of such CFCs are required to meet minimum detectability thresholds with JWST. (No potential white dwarf planetary target has yet been identified for such a search). \revision{\citet{haqq2022detectability} examined the detectability of CCl$_3$F and CCl$_2$F$_2$ on TRAPPIST-1e and found that present-day Earth abundances of these CFCs could show detectable spectral features with about 100\,hr of JWST time, assuming a low noise floor.} The launch of JWST \revision{thus represents} the first mission with the capability to detect atmospheric technosignatures, albeit with significant limitations.

Another example of an atmospheric technosignature is nitrogen dioxide (NO$_2$). The production of NO$_2$ on Earth today includes biogenic and anthropogenic sources, in addition to lightning. However, human generated NO$_{2}$ dominates by three times the amount from non-human sources \citep{Holmes2013}. Detecting high levels of NO$_2$ at levels above that of non-technological emissions found on Earth could be a sign that the planet may host active industrial processes \citep{Kopparapu2021}. NO$_2$ would be accessible in reflected light observation in the visible from LUVOIR and HabEx  space-based observatories for planets around Sun-like stars, and potentially observable in transit spectroscopy by JWST, Origins in the mid-infrared, and from ground-based ELTs for planets around M-dwarfs \citep{Kopparapu2021}. Other examples include CFCs infrared absorption bands between 7.8–15.3 $\mu$m \citep{rothman2009hitran} and sulfur hexafluoride (SF$_6$) absorption around 6-18 $\mu$m \citep{rothman2009hitran,sharpe2004gas}, which could potentially be accessible to JWST, Origins, and LIFE depending on their relative abundance and other factors. 

The search for atmospheric technosignatures will require ongoing effort to explore a range of plausible technosignatures and determine the specific atmospheric constituents that would be most detectable, prevalent, and unambiguous as a planetary technosignature. Although it may be difficult to predict the extent to which planetary technosignatures will be prevalent, it is worth emphasizing that the search for atmospheric technosignatures can be conducted alongside observations intended to detect atmospheric biosignatures or otherwise characterize planetary atmospheres with no additional marginal cost. 

\revision{Current, upcoming, and potential} future missions such as JWST, the IR/Optical/UV Surveyor recommended by Astro2020, LIFE, Nautilus, ground-based spectroscopy, and ELTs could be able to detect atmospheric technosignatures within restricted wavelength bands. Such a search for atmospheric technosignatures could be conducted alongside observations intended to detect atmospheric biosignatures or otherwise characterize planetary atmospheres with no additional marginal cost. Such observations could place upper limits on abundance of these gases, so long as they contain electronic or vibro-rotational features within the observed wavelength window.

\section{Artificial surface modifications}\label{sec:surface}

Another class of technosignatures are reflections or emissions from surface modifications on a planet. Some artificial surface modifications would require spatially-resolved observations to detect, while others may be detectable by planetary-integrated spectra. No significant searches for artificial surface modifications have yet been conducted.

\subsection{Reflection}

Artificial megastructures on the surface of a planet can be spatially resolved on exoplanets through inversions of reflected light curves \citep{berdyugina2019surface}. One example is subcontinent-size highly reflective artificial structures (e.g., to reject unwanted stellar light into space). Another is highly absorptive, photovoltaic-like structures (e.g., energy generators) in the near-planetary space above clouds. These structures can be spatially resolved and recognized because of their high contrast with respect to the natural environment and a possible regular substructure or shape. When reflected light curves are acquired in multiple passbands, spectral characteristics can help identify the nature of the structure. For example, \citet{Lingam2017} explored a possible spectroscopic signature of a planet-scale stellar energy collector, analogous to silicon photovoltaics producing a characteristic, detectable spectral “edge". Such technological spectral “edges” of alien global photovoltaic systems can be measured from spatially resolved multi-spectral albedo images of exoplanets \citep{berdyugina2019surface}. Identification of their constituent materials could then be possible in comparison to (photosynthetic or non-photosynthetic) biological edge features \citep[e.g.]{Berdyugina2016, seager2005vegetation, Hegde2015, Schwieterman2015} if their underlying structures are resolved. Existing capabilities for resolving surfaces of nearby exoplanets through inversions of optical light curves are ground-based photometry facilities and ELTs, while future facilities include LUVOIR, HabEx, and Nautilus.

\subsection{Heat islands}

Heat islands represent another possible technosignature on the surface of a planet. Kardashev Type I civilizations utilizing energy sources on the planetary scale may be ultimately forced to use exclusively photonic energy provided by the stellar radiation, to slow down an unavoidable global warming as a consequence of the second law of thermodynamics \citep{kuhn2015global}. If social aspects of advanced life are a universal feature, then civilization development may be clustered in favorable geographical areas where waste heat is constantly dumped into the environment raising its temperature to remotely detectable levels. Such ``heat islands'' are well identified with large cities in infrared Earth images taken from space \citep{Kim1992}. \citet{kuhn2015global} propose urban heat islands as a potentially observable technosignature for civilizations only slightly more advanced than ours (consuming $\sim$50 times more energy). A network of alien civilization heat islands may be resolved on the planetary surface and distinguished from environmental heat sources (e.g., volcanoes) through differential infrared measurements, e.g., at 5 $\mu$m and 10 $\mu$m. However, this observation would require an immense collecting area of approximately 70-m \citep{kuhn2015global}. No present capabilities exist for resolving surfaces of nearby exoplanets through inversions of near-infrared light curves, but such searches could be conducted by future ground-based facilities, ELTs, LIFE, and possibly Origins. 

\subsection{City lights}

One of the strongest spectroscopic technosignatures present on Earth’s nightside is the emission from nightside city lights, but on Earth this emission is relatively concentrated \citep{beatty2021}. Specifically, the nightside of the planet has an average surface flux of $0.035$ erg cm$^{-2}$ s$^{-1}$ from city light emission -- assuming 50\% cloud cover -- but the peak emission from New York City is 47 erg cm$^{-2}$ s$^{-1}$ sr$^{-1}$ in the few square kilometers around Times Square, while the peak emission from Toyko is 35 erg cm$^{-2}$ s$^{-1}$ sr$^{-1}$ in the Shinjuku area. Thus, while total surface flux from city lights on Earth is relatively low, this is primarily because the total area covered by cities on Earth is also low.

However, it may be that advanced civilizations on exoplanets have built cities over significantly more of their planets' surface. These more urbanized planets would have a higher nightside brightness from city lights, and be correspondingly easier to detect. An ecumenopolis, or planet-wide city, is the limiting case where an entire planet is completely covered by a single massive city. If we assume that the entire surface of an ecumenopolis has a constant surface intensity similar to city centers on Earth, of $40$ erg cm$^{-2}$ s$^{-1}$ sr$^{-1}$, and accounting for clouds, then an ecumenopolis will have a surface flux of roughly $63$ erg cm$^{-2}$ s$^{-1}$.

The dominant light source on Earth's night side is emission from street lights (or other area illumination lights), which reflect off nearby concrete and asphalt as seen from above. Modern street lights almost universally use high-pressure sodium (HPS) lamps of several hundred Watts, and though these are gradually being replaced by more energy efficient LEDs, we can make the simplifying assumption that all of Earth's street lights are HPS lamps. A typical HPS lamp emission spectrum has almost all of its emission concentrated between 550\,nm and 650\,nm, and a typical HPS bulb using 600W of input electrical power will emit approximately 130W of this input energy as light. This strong and narrow spectral feature of sodium emission lines can be readily distinguished from other sources of atmospheric emission.

The emission spectrum from city lights on Earth as observed from space is also effected by the albedo spectrum of concrete, and the transmissivity of the atmosphere. For the former, most of the emission we would see from an exoplanet is caused by the light emitted by downward facing artificial lights reflecting back up, off nearby surfaces. On the Earth, these surfaces are predominately concrete buildings and roads, which has a reflectance spectrum that is effectively flat over the wavelengths of interest. The transmission of the atmosphere does have a significant effect on the emission spectrum, however, and needs to be included in any calculations.

\cite{beatty2021} investigates the detectability of city lights under the above assumptions for both LUVOIR and HabEx using 100 hour observations. While Earth-like urbanization fractions would not be detectable by any of the LUVOIR or HabEx architectures, the detection of near-Earth urbanization fractions would be possible. \revision{In particular, LUVOIR A imaging of Proxima Centauri b (assuming it is a habitable rocky planet hosting a civilization on its surface) would be capable of detecting city lights from an urbanization fraction of 0.006, or 0.6\% . This is about twelve times the urbanization fraction on Earth.}

\revision{An ecumenopolis, or planet-wide city, would be detectable around roughly 30 to 50 nearby stars by both LUVOIR and HabEx. A general survey of these systems would place a $1\,\sigma$ upper limit of $\lesssim2\%$ to $\lesssim4\%$, and a $3\,\sigma$ upper limit $\lesssim10\%$ to $\lesssim15\%$, on the frequency of ecumenopolis planets in the Solar neighborhood assuming no detections.}

\section{Optical beacons \& Optical SETI}\label{sec:optical}

\revision{Shortly after the search for radio technosignatures was first proposed by \citet{cocconi1959searching}, a similar idea of searching for continuous wave optical laser signals was suggested by \citet{schwartz1961interstellar} as another means of interstellar communication. Optical beacons provide an additional range of wavelengths to search in, where interstellar signals could be encoded and transmitted with far more information than is possible with radio \citep{ross1965search,bhathal2000case}. Continuous wave observations, however, require searches at specific monochromatic frequencies instead of a range \citep{kingsley2001optical} and pulsed wave (temporal) signals were soon proposed as an alternative \citep{ross1965search, shvartsman1977mania}. A pulsed wave signal is far less energy intensive, able to transmit across a broadband of frequencies and can produce visible light bursts brighter than our sun, thus creating an optical signal that would be easily detectable against background interference \citep{howard2000optical}.}

\revision{Although most previous searches for intentional signaling or communication have focused on radio beacons, a number of searches for optical technosignatures have also been conducted using ground-based facilities. Beginning in the 1970s and through until the end of the 20th century, optical SETI was conducted extensively with the MANIA (Multichannel Analysis Nanosecond Intensity Alterations) experiment, which searched for pulsed wave signals using a 6 m ground-based telescope \citep{shvartsman1977mania}. Despite these efforts MANIA produced null results. After MANIA, various continuous and pulsed laser searches were conducted including preparations for a 25 cm Columbus Optical SETI Observatory (COSETI) \citep{10.1117/12.243436}, thousands of observations of F and M stars with the Lick Observatory \citep{stone2005lick}, surveys of southern circumpolar stars and globular clusters \citep{bhathal2001optical}, and all-sky surveys to place limits on the density of pulsed laser signals in the galaxy \citep{howard2007initial}. More recently \citet{tellis2015search,tellis2017search} studied high-resolution data from Keck observations, looking for evidence of continuous wave laser emission lines from nonnatural sources. And archival observations made by the Very Energetic Radiation Imaging Telescope Array System (VERITAS) were used to look for pulsed optical beacons around Boyajian's star \citep{abeysekara2016search}. To date none of these endeavours have found evidence of optical extra-terrestrial signals but future work is continuing the optical SETI mission.}

\revision{The Pulsed All-sky Near-infrared Optical SETI (PANOSETI) observatory is a design for a dedicated facility to perform optical and near-infrared surveys for nanosecond pulses across the entire northern hemisphere \citep{2019BAAS...51g.264W}. And the Breakthrough Listen project is collaborating with VERITAS to search for pulsed optical beacons \citep{gajjar2019breakthrough}. Many ground-based facilities could be used to place limits on the prevalence of optical beacons, while larger facilities such as Rubin, RST, and ELTs---and perhaps ARIEL and EarthFinder---could provide even stronger detectability constraints. The next-generation space-based observatories the Infrared/Optical/Ultraviolet space telescope and Nautilus---and possibly HST and JWST---could provide further constraints on the presence of optical beacons in exoplanet systems. \citet{vides2019model} have produced predictions for the Gemini planet Imager (GPI) and RST. Their work predicts observational capabilities for a continuous laser signal for both observatories and determine that 24 kW (GPI) and 7.3 W (RST) signals would be detectable from the $\tau$ Ceti system. An important note is that the 7.3 W predicted signal can be detected from within $\tau$ Ceti's habitable zone. This suggests future-generation large-aperture space-based observatories such as the IR/O/UV telescope maybe be able to detect low-powered optical signals from terrestrial exoplanets that are characterizable and within their host stars habitable zone.}

\section{System Megastructures}\label{sec:megastructures}

Megastructures and other space engineering projects represent another class of detectable technosignatures. Such structures could be constructed for a variety of regions and could be detectable through transit observations or from observations of thermal infrared excesses. Dysonian megastructures are one example but there are a number of other potential transiting space engineering technosignatures that current and future missions could observe.

\subsection{Transiting megastructures}

Dyson spheres (or Dyson swarms) represent one of the first megastructures to be suggested in the history of technosignature science and remain a possibility for detecting evidence of an energy-intensive extraterrestrial civilization \citep{dyson60,2020SerAJ.200....1W}. Dysonian megastructures \revision{refer to the idea that a technological civilization could construct} large energy collectors in orbit around the host star in order to maximize the energy utilization of the system. Relevant technosignatures could be any evidence of these massive objects, which could potentially be detected using a combination of ground and space based observatories. Current and retired space observatories have already shown the potential for their application in these searches, such as the anomalous transit events detected by Kepler for the KIC 8462852 system (also known as Boyajian's Star or Tabby’s Star \revision{\citep{Boyajian2016}}). Even though evidence now points to a more natural explanation for the dimming of KIC 8462852 \revision{such as dust, comets, trojan asteroids, and/or planets with rings \citep[e.g.,][]{bodman2016kic, wright2016families, Neslusan2017, Ballesteros2018}} the observations demonstrated the value of current data and archival data in the search for evidence of megastructures. \revision{Ongoing work is being done to identify similar anomalous signals that may have implications for the detection of megastructure technosignatures \citep[e.g.,][]{giles2019systematic,Chakraborty2020,Schmidt2021}.}

Megastructures could also exist as orbiting structures around an exoplanet, which could also be detectable as transit events. With current capabilities, such devices would need to be very large, possibly of planetary size, or consist of a swarm of smaller objects, such as a large number of satellites. A number of possible scenarios that would leave an imprint on a stellar light curve have been proposed in the literature \cite[e.g.][]{arnold2005transit, Wright2015}. One example is the construction of a starshade at the L1 Lagrange point of the star-planet system as a way of mitigating problems of climate change \citet{gaidos2017transit}. A related possibility is the use of a fleet of orbiting mirrors in orbit around a synchronously rotating planet around a low-mass star as a way of artificially warming the surface on the planet's cold night side \citet{korpela2015modeling}. Another possibility is the detection of a large population of geosynchronous devices, known as a Clarke exobelt, which could remain detectable for millions of years and even outlast the host civilization \citet{socas2018possible}. Other artificial satellites can stay in their orbits by using solar radiation to oppose the force of gravity (known as Quasites \citep{kipping2019transiting} and Statites \citep{forward1993statite}) and could exhibit unique transit signatures. An extraterrestrial civilization could conceivably even use artificial transiting structures as a means of interstellar communication \citep[e.g.][]{arnold2005transit,korpela2008session}. This list is by no means exhaustive but is intended to illustrate the possibilities for planetary scale megastructures.

The difficulty in conducting a search for transiting megastructures is distinguishing celestial objects, such as moons or rings, from built structures or devices. Some of the scenarios described above would produce signatures that may be relatively straightforward to recognize as a technosignature, while others may be more ambiguous even if observed. New instrumentation for exoplanet transit observation is opening possibilities for conducting such searches. For example, a number of researchers are engaged in a search for \revision{exomoons} \citep[e.g.][]{kipping2013hunt, Hippke2015exomoon}. \revision{Two exomoon candidates include the Neptune-sized Kepler-1625 b-i \citep{Teachey2018}, although its discovery may be spurious \citep{Kreidberg2019}, and the recently discovered 2.6 Earth-radii candidate Kepler-1708 b-i \citep{Kipping2022}}. There is also significant scientific interest in the discovery of exorings \citep[e.g.][]{heller2018detecting}. All outer planets in the solar system (and some minor bodies) have rings but no proper ring system has yet been identified around an exoplanet. Such search efforts could similarly detect the presence of megastructures in exoplanetary systems, \revision{as could searches for trojan asteroids in exoplanetary systems \citep{hippke2015statistical}.}  

Many current and future facilities could place limits on the presence of stellar and planetary megastructures in exoplanetary systems. Studies have been performed to see if future observatories will be able to participate in searches for Dysonian megastructures, such as for the Gaia mission \citep{zackrisson2018seti}. The authors noted that the best candidate they studied may have had its distance measurements affected by a natural object (a white dwarf), but further data releases should improve this technosignature detection technique. Numerous ground-based and space-based facilities could be used to place constraints on the prevalence of transiting Dysonian megastructures, such as Kepler, TESS, JWST, HST, RST, PLATO, LUVOIR, HabEx, and Nautilis. Other facilities that might be able to place some limits on megastructures in exoplanetary systems include CHEOPS, Rubin, ARIEL, ELSs, EarthFinder, PANOSETI, and Origins. 

\subsection{Waste heat}

Megastructures could also be identified by an excess of infrared radiation emission that results as waste. The original suggestion to search for megastructures by \citet{dyson60} focused on identifying artificial sources of infrared radiation at the scale of a planetary system or even an entire galaxy. Constraints on the prevalence of megastructures in exoplanetary systems can be placed by ground-based and space-based spectroscopy facilities at infrared wavelengths, while any megastructure candidates identified through transits can be subsequently searched for evidence of anomalous infrared radiation emissions. Facilities such as NeoWISE and Spitzer are ideal for searching for such infrared excesses, while JWST, ELTs, and Origins could also be capable of providing such constraints.

The same principle can apply to a planet, as megastructures in a planetary orbit could also be identified, at least in principle, by any anomalous infrared radiation they emit as waste. Methods for identifying planetary infrared excess radiation are being developed for the characterization of non-transiting planets \citep{stevenson2020new}, which primarily focuses on warm and closely orbiting planets but theoretically could be used to place constraints on the presence of planetary-scale megastructures in some systems.

\section{Interstellar spaceflight}\label{sec:spaceflight}

A final possible technosignature to consider is the emission from a spacecraft’s drive while it is under acceleration. The time spent under acceleration with an ``active'' drive is generally very short compared to the total mission time for present-day space missions, which would make detecting present-day spacecraft drives difficult. But there are some present-day propulsion systems (e.g., ion engines) which are designed around continuous acceleration. In the future, it is possible to imagine the sort of continuously operating drives, but based on higher energy processes such as fission, fusion, or antimatter \citep[e.g.][]{zubrin1995detection}. 

One possible realization of a fusion propulsion system would be a system based on D-$^3$He fusion. The primary emitters from such a D-$^3$He propulsion system would be the central fusion drive core and the fusion exhaust. The former can be approximated as a sphere of optically thick gas 50\,cm in radius, emitting as a blackbody at a temperature of 9x10$^8$ K. The drive exhaust would primarily be composed of fully-ionised $^4$He plasma, which would emit the $^4$He recombination spectrum as the plasma ions cooled and recombined. Such an exhaust could have a high density of  $~\sim$10$^{18}$ cm$^-1$ as it leaves the drive core and would primarily emit for the first $\sim$1-10 km behind the spacecraft. \citet{zubrin1995detection} first examined the detectability of interstellar spacecraft and concluded that gamma radiation emitted by such systems would probably be undetectable, but visible light signatures could be detectable for up to hundreds of light years and low frequency radio emissions to thousands of light years. The detection of emission spectrum features from such a spacecraft would be a technosignature and could conceivably be detectable by large ground-based and space-based spectroscopy facilities; however, further work is needed to determine detectability limits for any current or future facilities.

\revision{Laser propulsion remains another possibility that could be detected at interstellar distances. Specifically, the Breakthrough Starshot initiative has begun the concept design for a system of laser-propelled nanocraft that would be sent toward the Alpha Centauri system \citep{parkin2018breakthrough}. The Breakthrough Starshot system would require a ground-based laser of power $\sim$100\,GW or greater. Such a propulsion system could be detectable by ground- and space-based optical facilities if a similar extraterrestrial laser were directed toward or through the Solar System.}

\section{Discussion}

The overview of technosignatures provided in this paper is intended to illustrate the breadth of possibilities constraining the presence of extraterrestrial technology through the use of ultraviolet, optical, and infrared observing facilities. The references in this paper are by no means exhaustive, and the list of possible exoplanetary technosignatures may include others not discussed here. Nevertheless, the previous sections of this paper have demonstrated a theoretical basis for considering a variety of exoplanetary technosignatures as well as possible obervational methods for placing constraints on the prevalence of such technosignatures. The science technosignatures remains in an early stage compared to the science of understanding non-technological biosignatures, so much more research is needed to understand the detectability limits for all of the technosignatures discussed in this paper.

The search for technosignatures can also advance using current and future mission\revision{s and facilities} without the need for significant additional funding. Astronomers engaged in observing or data analysis with any of the missions \revision{or facility types} listed in Fig. \ref{fig:currentfuturemissions} could include consideration of technosignatures as part of their research, with commensal observing possible in many cases. Future missions can likewise include the search for technosignatures as an additional science justification, without needing to dedicate any additional resources or design considerations. Many technosignatures searches are already conducted by using commensal or archived data from other missions, so emphasizing the connection between general-purpose astronomical missions and the search for technosignatures can help to advance the science without incurring significant costs. \revision{Importantly, the examples given in Fig. \ref{fig:currentfuturemissions} are meant to be illustrative rather than exhaustive, and focus on missions or facilities with the capability of characterizing terrestrial planets. There are a wide variety of ongoing or planned NASA\footnote{\url{https://www.nasa.gov/content/universe-missions-list}} and ESA\footnote{\url{https://www.esa.int/ESA/Our_Missions}} missions that could potentially support commensal technosignature searches. Many opportunities exist for identifying and leveraging these missions (or ground-based facilities) to search for technosignatures.}

For more distant mission planning, technosignature science may eventually serve as a primary science goal that drives some features of design. Some ground-based facilities are dedicated to, or optimized for, the search for technosignatures today, although these are primarily searches for narrow-band radio signals or optical beacons. Current mission concepts for searching exoplanetary systems for signs of life are driven by biosignature science, and it remains conceivable that technosignature science could advance to the stage at which mission design is driven by specific technosignature searches. Ideas for such dedicated technosignature mission concepts were proposed by \citep{socas2021concepts}. Further technosignature research will continue to develop a broad set of possibilities for detecting technosignatures, as well as false positives, in an effort to eventually prioritize candidate technosignatures for actual searches. 

\revision{Additional advances in technosignature research are also possible through advances in Earth observing. As Earth is the only known planet to host life, as well as technology, observations of Earth can serve as proxies for exoplanet observations. This idea of studying Earth as an example of an inhabited exoplanet includes examining Earth's spectral biosignature \citep[e.g.,][]{robinson2018earth,jiang2018using,klindvzic2021loupe} as well as observing Earth's radio spectrum as reflected from the moon \citep[e.g.,][]{sullivan1978eavesdropping,mckinley2012low}. Future Earth observing can provide opportunities to further constrain the detectability of Earth's technosignatures.}

\revision{Advances in machine learning and other computational methods also provide opportunities for conducting technosignature searches using new and existing data. Machine learning methods have shown promise in the search for radio technosignatures as a way to search large datasets \citep[e.g.,][]{ma2022first} or to filter out radio interference \citep[][]{pinchuk2022machine}. Machine learning has also been used to search for anomalies in data from the \textit{Kepler} mission \citep{giles2019systematic,giles2020density}. Further development of computational methods can help to place further constraints on the prevalence of technosignatures without necessarily conducting new observations.}

The summary of mission capabilities shown in Fig. \ref{fig:currentfuturemissions} highlights some important technology gaps in current and future capabilities in the search for technosignatures. The ability to detect infrared technosignatures is limited to a handful of facilities, with the capabilities of JWST, ELTs, and mission concepts like Origins as the most likely to be able to conduct such searches in nearby systems. Many current and near-future facilities are beginning to show promise in the search for transiting megastructures as well as optical beacons, which represents an encouraging development in the search for technosignatures by expanding beyond radio wavelengths. However, searching for atmospheric or surface technosignatures is primarily a feature of the next generation of space telescopes and ELTs. \revision{Such facilities could conceivably also be used to detect evidence} of interstellar flight in exoplanetary systems, so further study of this technosignature class in particular would be useful for determining detectability limits with future facilities.

Finally, it is worth emphasizing that null results are valuable in the search for technosignatures. Any astronomical survey that is able to place a detection limit on one or more technosignature helps to constrain our knowledge about the prevalence of such features in the galaxy. Such null results can be possible with existing archival data and may not even require much additional time on the part of the researcher. Current and future observing facilities can therefore play a valuable role in constraining the prevalence of technosignatures that are within detectability limits, which will help to narrow the scope of the search.

\section{Conclusions}

Many current and future missions are \revision{suited for advancing the search for technosignatures.} The purpose of this paper has been to serve as a resource for the growing community of technosignatures researchers by describing the capabilities of present-day technology to constrain the search for technosignatures. Many searches for technosignatures could be conducted with existing data or through commensal observing with other programs, which represent opportunities to advance technosignature science without the need to invest in new major resources. The broad range of technosignatures across the full electromagnetic spectrum should also be included as ancillary science justifications for mission proposals, as many ongoing exoplanet science efforts (as well as solar system science) could help to constrain the presence of certain technosignatures.

One of the prevailing themes from the TechnoClimes 2020 workshop discussions was the need to engage the broader astronomical community in thinking seriously about the possibility of detecting technosignatures. The concepts discussed in this paper and elsewhere in the literature provide hypothesis-driven examples of technosignatures that could be detected, but the first detection of a technosignature could also be an anomalous discovery that was not anticipated by any of these examples. In either case, support for technosignature science through cooperation among interdisciplinary teams of astronomers (and adequate support from funding agencies) will be essential to advance the search. The tools to find technosignatures may already be available, but it will require a community-wide effort to start looking.

\section*{Acknowledgments}

This study resulted from the TechnoClimes workshop (August 3-7, 2020, technoclimes.org), which was supported by the NASA Exobiology program under award 80NSSC20K1109. The authors thank James Benford, Hector Socas-Navarro, and Jason Wright for providing comments on this manuscript. We also thank three anonymous referees whose comments allowed us to improve the manuscript. 

RK acknowledges support from the GSFC Sellers Exoplanet Environments Collaboration (SEEC), which is supported by NASA’s Planetary Science Division’s Research Program. This work was performed as part of NASA’s Virtual Planetary Laboratory, supported by the National Aeronautics and Space Administration through the NASA Astrobiology Institute under solicitation NNH12ZDA002C and Cooperative Agreement Number NNA13AA93A, and by the NASA Astrobiology Program under grant 80NSSC18K0829 as part of the Nexus for Exoplanet System Science (NExSS) research coordination network. HSN acknowledges support  from the Spanish Ministerio de Ciencia, Innovaci\'on y Universidades through project PGC2018-102108-B-I00 and FEDER funds. EWS acknowledges support from the NASA Astrobiology Program under Cooperative Agreement Number NNA15BB03A and grant number 80NSSC18K0829. SS acknowledges the SETI Forward Award from SETI Institute. Any opinions, findings, and conclusions or recommendations expressed in this material are those of the authors and do not necessarily reflect the views of their employers, NASA, or any other sponsoring organization.

\section*{Data Statement}
No new data were generated or analysed in support of this research.

\bibliography{main}{}

\end{document}